\documentclass[%
reprint,
longbibliography,
amsmath,amssymb,
aps,
]{revtex4-1}

\usepackage{color}
\usepackage{textcomp}
\usepackage{hyperref}
\usepackage{graphicx}
\usepackage{dcolumn}
\usepackage{bm}

\usepackage{tikz}

\usepackage{tcolorbox}

\usepackage[tikz]{bclogo}

\begin{document}
	
	\preprint{APS/123-QED}
	
	\title{Community detection in network neuroscience}
	
	\author{Richard F. Betzel$^{1-4}$}
	\email{rbetzel @ indiana.edu}
	
	\affiliation{
		$^1$Department of Psychological and Brain Sciences, $^2$Cognitive Science Program, $^3$Program in Neuroscience, $^4$Network Science Institute, Indiana University, Bloomington, IN 47405
	}

	
	\date{\today}
	\begin{abstract}
		
		Many real-world networks, including nervous systems, exhibit meso-scale structure. This means that their elements can be grouped into meaningful sub-networks. In general, these sub-networks are unknown ahead of time and must be ``discovered'' algorithmically using community detection methods. In this article, we review evidence that nervous systems exhibit meso-scale structure in the form of communities, clusters, and modules. We also provide a set of guidelines to assist users in applying community detection methods to their own network data. These guidelines focus on the method of modularity maximization but, in many cases, are general and applicable to other techniques.
		
	\end{abstract}
	
	\maketitle
	\section*{Introduction}
	
\begin{figure*}[t]
		\centering
		\includegraphics[width=1\textwidth]{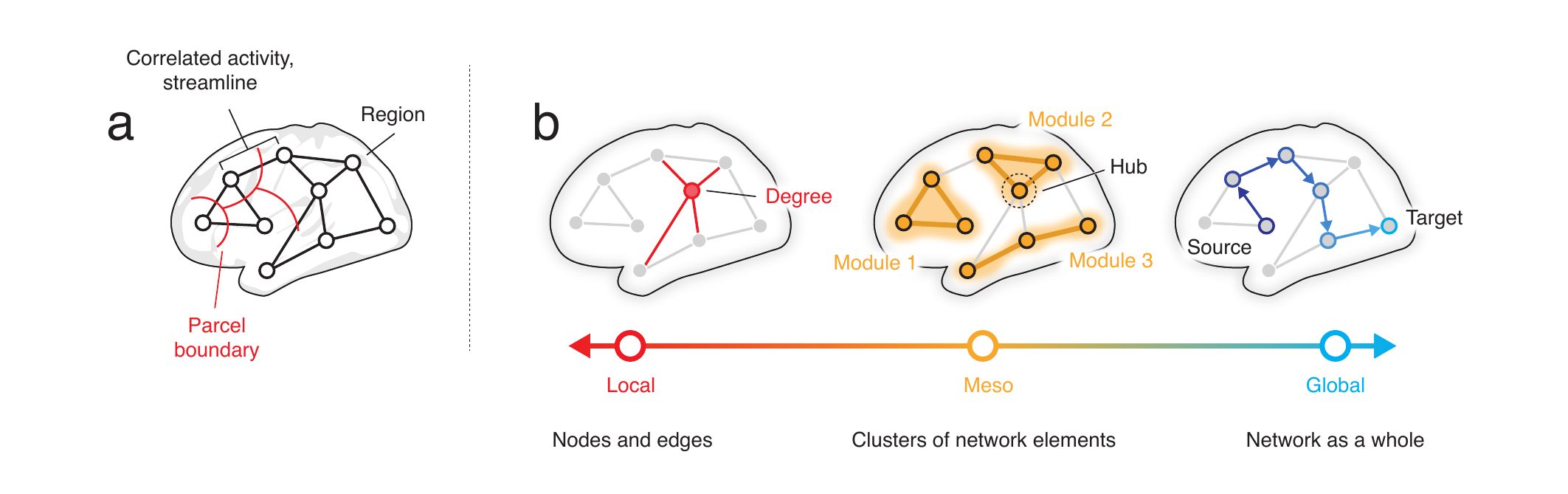}
		\caption{\textbf{Scales of network analysis.} (\emph{a}) Brain networks are usually defined as collections of neural elements (cells, populations, regions) linked to one another \emph{via} structural or functional connections. (\emph{b}) Analysis of brain networks can be carried out at the local and global scales, which focus on properties of individual network elements (nodes and edges) or the network as a whole, respectively. Situated between these extremes is a \emph{meso-scale} that focuses on sub-networks, i.e. groups and clusters of nodes or edges.} \label{networks+scales}
	\end{figure*}

	The human brain is a complex network, comprised of neurons, neuronal populations, and brain areas linked to one another by structural and functional connections \cite{sporns2005human, park2013structural}. This network can be expressed as a graph of nodes and edges and analyzed using methods from network science \cite{bullmore2009complex, bassett2017network} (Fig.~\ref{networks+scales}a). Analyses of human brain networks have begun to reveal the key organizational principles of brain networks, including small-world architecture \cite{achard2006resilient, sporns2004small}, presence of hubs and rich clubs \cite{hagmann2008mapping, van2011rich}, and cost-efficient wiring diagrams \cite{bassett2010efficient}.
	
	Many network analyses use graph-theoretic measures to describe features of a brain network \cite{rubinov2010complex}. Some measures focus on properties of the brain at a \emph{local} scale (Fig.~\ref{networks+scales}b), characterizing the number of connections a node makes, its centrality to the network, or the extent to which its neighbors are also connected to one another. Other measures describe \emph{global} properties of the network as a whole (Fig.~\ref{networks+scales}b), characterizing its paths and cycles to reveal how information flows through the system.
	
\begin{figure*}[t]
		\centering
		\includegraphics[width=1\textwidth]{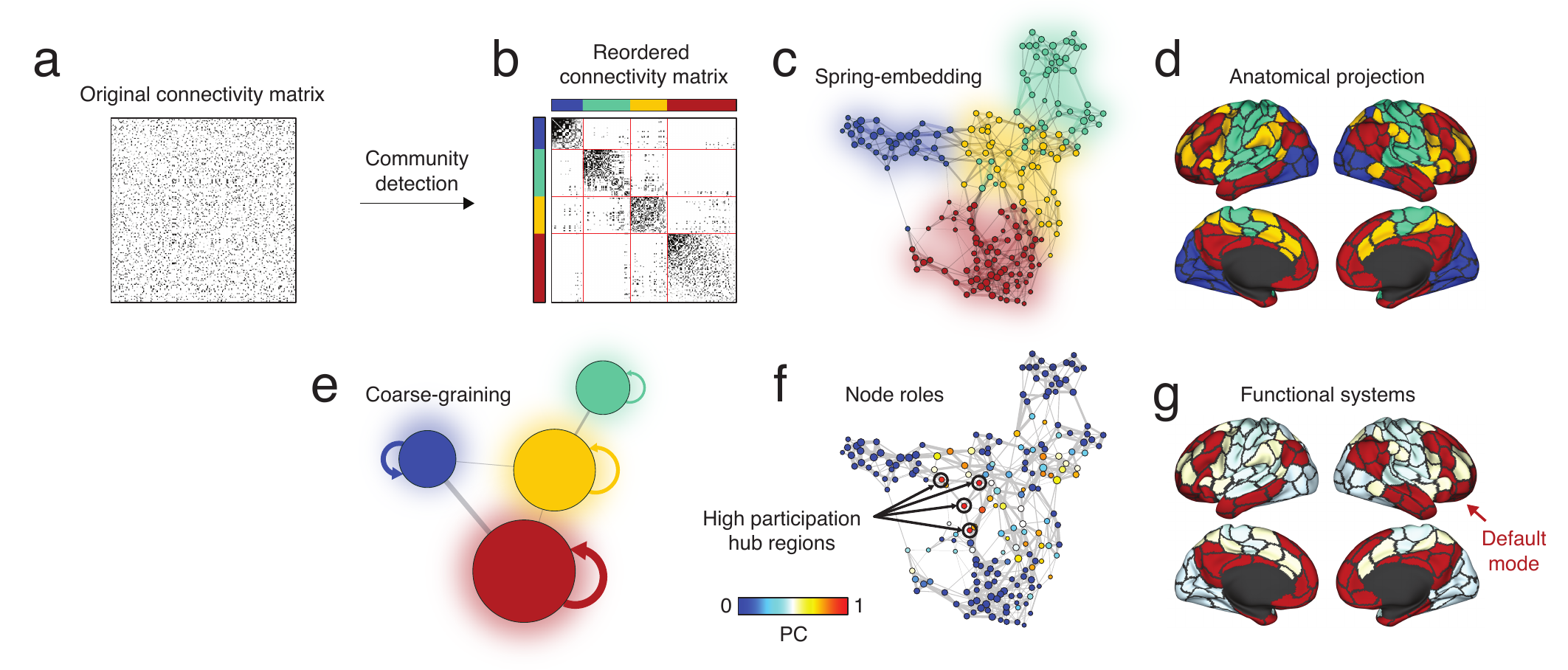}
		\caption{\textbf{Schematic illustrating why community detection is necessary and what can be gained from it.} (\emph{a}) Community detection usually starts with a network's connectivity matrix. Initially, no community labels are known and its rows and columns are in an arbitrary order, making it difficult to establish whether it has communities. (\emph{b}) Community detection algorithms can be applied to this matrix to reveal its meso-scale structure. Here, we show the original matrix with its rows and columns reordered to highlight the ``block diagonal'' structure of assortative communities. We can visualize these communities through a spring-embedding (\emph{c}) or by projecting node's community labels onto brain anatomy (\emph{d}). (\emph{e}) The advantage of this approach is that it allows us effectively reduce the dimensionality and complexity of a network by treating its communities rather than its nodes as the units of interest. (\emph{f}) With community labels, we can classify nodes' roles by describing how each node's links are concentrated within or distributed across communities. (\emph{g}) In the case of brain networks, community labels can be used to identify functionally-related groups of nodes, e.g. communities that recapitulate known intrinsic connectivity networks like the default mode.} \label{why+community+detection}
	\end{figure*}

	Both local and global scales represent extremes. Situated between the two is a rich topological \emph{mesoscale} that focuses on groups of nodes that are sometimes called sub-networks, sub-graphs, communities, or modules \cite{sporns2016modular, meunier2010modular} (Fig.~\ref{networks+scales}b). A network's community structure is of broad interest scientifically (Fig.~\ref{why+community+detection}). It represents a condensed, coarse-grained view of a network by shifting focus onto clusters rather than individual elements (Fig.~\ref{why+community+detection}e). It can even be used to further characterize the functional roles of individual nodes (Fig.~\ref{why+community+detection}f), for example as connector hubs whose connections form bridges between two communities, or to discover groups of nodes and edges that likely participate in similar network functions (Fig.~\ref{why+community+detection}g). 
	
	Importantly, in the context of brain networks, the methods used to study a network's community structure can be applied profitably to both structural \cite{baum2017modular} and functional connectivity \cite{power2011functional} (and even used to explicitly link the two modalities \cite{betzel2013multi}). Brain network communities also represent a potentially powerful and flexible substrate for biomarker generation \cite{xia2020multi, alexander2010disrupted} and to help discovering links between brain, behavior, and disease \cite{bertolero2018mechanistic} .
    
	While a network's community structure can offer insight into its organization and function, it is usually unknown ahead of time and cannot be inferred from visual inspection alone. Rather, a network's communities must be detected algorithmically through a process known as \emph{community detection} \cite{fortunato2010community, fortunato2016community}.
	
	This article focuses on the topic of community detection from a practical, user-oriented perspective. However, its aim is not to provide a definitive roadmap for community detection. Rather, it is to convince the reader that community detection is a fundamentally difficult process that is guided by the user, who, even for the ``simplest'' algorithms, must make a series of non-trivial decisions. On one hand, these decisions present challenges to the user. This is especially true for researchers new to network neuroscience who might be running community detection for the first time. On the other hand, these decisions afford the user some agency, making community detection a flexible framework that can be directed toward a wide range of research questions.

	


	\section*{Background}
	
	Communities refer to meaningful sub-networks -- also referred as modules or clusters -- embedded within a larger global network \cite{newman2012communities}. Although sub-networks can form core-periphery and multi-partite structures, they are usually assumed to be ``assortative,'' meaning that the nodes in a given community tend to be strongly connected to one another but weakly connected to nodes in other communities. This type of structure helps support autonomous, specialized function \cite{espinosa2010specialization}, buffer the impact of damage away from the rest of the network \cite{nematzadeh2014optimal}, reduce the material and metabolic cost of wiring \cite{bassett2010efficient}, and support separation of dynamic timescales \cite{pan2009modularity}. Assortative communities also promote a network's evolvability \cite{kirschner1998evolvability}, allowing individual communities to explore phenotypic variation without compromising the broader functionality of the network as a whole.
	
	Indeed, modular structure has been observed in empirical brain network data, both in structural \cite{hagmann2008mapping} and functional \cite{power2011functional} brain networks reconstructed from virtually all imaging and recording modalities and across every spatial scale \cite{glomb2019using, martinet2020robust, shih2015connectomics, mann2017whole, bruno2015modular}. From the whole-brain neuronal connectomes of \emph{C. elegans} \cite{jarrell2012connectome}, zebrafish \cite{betzel2020organizing, hadjiabadi2020higher}, and drosophila \cite{mann2017whole}, to populations of neurons \cite{shimono2015functional, betzel2019stability, perez2020long}, to fast electrophysiological recordings \cite{martinet2020robust, betzel2019structural}, to large-scale human brain networks \cite{yeo2011organization} -- wherever neuroscientists search for communities, communities tend to appear.
	
	The analysis of community structure has had its greatest impact in the neuroscientific community at the macroscale, where brain networks are reconstructed from functional and diffusion MRI \cite{craddock2013imaging}. Early highly influential studies applied community detection algorithms to whole-brain resting-state functional networks to reveal non-overlapping modules whose boundaries, surprisingly, were closely aligned with previously delineated brain systems and task-evoked patterns of activity \cite{yeo2011organization, power2011functional}. These modules, which bore a striking resemblance to maps of correlated activity recovered from independent component analyses \cite{damoiseaux2006consistent}, mapped to distinct cognitive domains \cite{smith2009correspondence}, and are recapitulated in meta-analytic networks of task co-activations \cite{crossley2013cognitive}. These observations suggest that, even in the absence of explicit task instructions, the brain's intrinsic architecture is modular and subtends cognitive function.
	
	Similar findings have been reported in networks of anatomical connectivity, although the correspondence of functional and anatomical modules is inexact and not one-to-one. For instance, in Hagmann et al \cite{hagmann2008mapping}, the authors reported six modules: two located along the midline and four distributed in either hemisphere. In contrast to functional modules, which are spatially distributed and oftentimes comprised of multiple extended sub-modules, anatomical modules tend to be spatially contiguous, reflecting the effect of material and metabolic constraints on the brain to reduce its wiring cost \cite{stiso2018spatial}, a tendency that has been replicated in recent studies \cite{mivsic2015cooperative, baum2017modular}.
	
	Collectively, evidence abounds that nervous systems are modular. These modules span multiple spatial scales and network modalities, revealing the brain's rich multi-scale and hierarchical communities.
	
	
	\section*{Community detection: Methods description}
	
	In most applications, a network's community structure is unknown ahead of time and cannot be inferred based on visual inspection alone. Instead, we use algorithms and heuristics to partition nodes and edges into sub-networks, called communities, modules, or clusters. This enterprise is referred to as ``community detection'' and involves many steps \cite{fortunato2010community, fortunato2016community}. Because the space of community detection methods is paralyzingly massive, it is impossible to discuss every method and contingency as part of this article. In the following sections we break down some of the common steps for performing community detection and decisions that a user has to make, focused through the lens of the \emph{modularity heuristic}.
	
	\subsection*{Choosing an algorithm}

	\begin{tcolorbox}
		
	\textbf{Box 1}. Different algorithms define communities in different ways and can yield dissimilar estimates of your network's community structure. For instance, modularity maximization is designed to detect internally dense and external sparse assortative communities. Stochastic blockmodels can detect more general classes of communities. Care should be taken in selecting the algorithm. 
		
	\end{tcolorbox}
	
	The first step in community detection is to choose an algorithm. This step is critical -- every algorithm operates according to a different definition of what it means for a group of nodes or edges to form a community. Additionally, some algorithms require first transforming your network in some way, e.g. by discarding or rectifying negative connection weights. More generally, algorithms encode a set of assumptions about how your network was generated and what its communities look like. Consequently, the choice of algorithm has profound impact on the character of the detected communities.
	
	Methods like Infomap \cite{rosvall2008maps}, for instance, define communities to be collections of nodes that constrain the probabilistic flow of a random walker moving over the network. As a result, Infomap detects communities that are ``assortative,'' with connection density concentrated within communities compared to between. However, because Infomap is based on random walks which are only defined for networks with positively-weighted edges, all negative weights must be first thresholded.
	
	In contrast, stochastic blockmodels \cite{wasserman1994advances, karrer2011stochastic} use statistical inference to recover the parameters (including community labels) of the generative model that gave rise to an observed network. In the classical blockmodel, every node belongs to a community and edges are generated stochastically based on some underlying distribution that depends only on the community to which an edge's stub nodes belong. Consequently, blockmodels can detect more general classes of communities; including core-periphery and disassortative configurations (where between-community connection density exceeds the within-community density) \cite{betzel2018diversity, faskowitz2018weighted}.
	
	Both Infomap and the stochastic blockmodel assign nodes to non-overlapping communities, meaning that each node is assigned to one community and one community only. However, in many real-world networks nodes belong to multiple communities. Methods like link communities \cite{ahn2010link}, line graph clustering \cite{evans2009line}, and clique percolation \cite{palla2005uncovering} are capable to detecting overlapping communities although, as before, the precise definition of what constitutes a community varies between these methods.
	
	\subsection*{The modularity heuristic}
	
	\begin{tcolorbox}
		
		\textbf{Box 2}. Modularity maximization is a widely used community detection method. It defines communities as groups of nodes that are more densely connected in your observed network than you would expect by chance. The $Q$ value associated with a partition -- its modularity -- is a measure of its quality. Larger values usually imply higher quality partitions.
		
	\end{tcolorbox}
	
 	Among the most widely used community detection algorithms is modularity maximization \cite{newman2004finding}. Modularity maximization operates according to an eminently simple principle: compare what you see with what you expect. More specifically, modularity defines communities to be groups of nodes more densely connected to one another than would be expected had the network been generated by a random null model. This intuition is formalized by the modularity heuristic, $Q$, which serves as an objective function for quantifying the quality of any partition and is calculated as:
	
	\begin{equation}
	Q = \sum_{ij} B_{ij} \delta(\sigma_i,\sigma_j).
	\end{equation}
	
	In this expression, $B_{ij}$ is an element taken from the \emph{modularity matrix}, which is computed as the difference in each connection's observed weight, $W_{ij}$ (Fig.~\ref{null+models}a), with the weight we would expect it to have under the null model, $P_{ij}$ (Fig.~\ref{null+models}b), such that $B_{ij} = W_{ij} - P_{ij}$ (Fig.~\ref{null+models}c). The variable $\sigma_i \in \{  1 , \ldots , K \}$ encodes the community to which node $i$ is assigned, and $\delta(\sigma_i,\sigma_j)$ is the Kronecker delta function and is equal to 1 when its arguments are identical and zero otherwise. Although the summation is carried out over all pairs of nodes, the delta function ensures that the only node pairs that increase or decrease $Q$ are those that fall within the same community. In general, partitions corresponding to larger values of $Q$ are considered ``better'' than those with smaller values.
	
	\subsubsection*{Null models}
	
	\begin{tcolorbox}
		
		\textbf{Box 3}. Modularity, $Q$, is calculated by comparing your real network with the network you would expect given some a null connectivity model. The choice of null model is critical and should be motivated by research question and properties of your network, e.g. structural \emph{versus} functional connectivity or signed \emph{versus} positive-only connection weights. In general, different null connectivity models will yield dissimilar estimates of community structure.
		
	\end{tcolorbox}
	
	\begin{figure*}[t]
		\centering
		\includegraphics[width=1\textwidth]{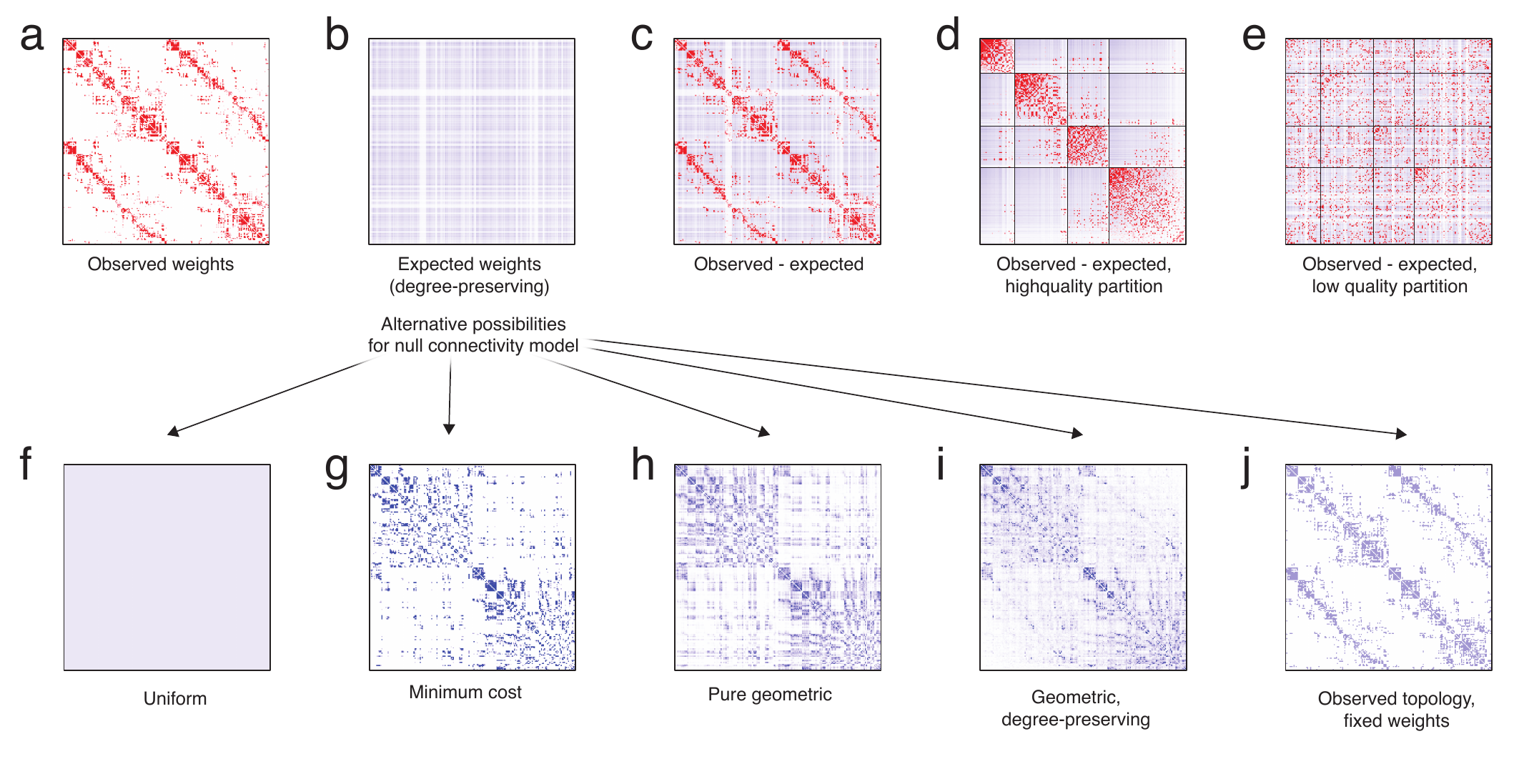}
		\caption{\textbf{Modularity maximization and choice of null model.} Modularity maximization works by comparing weights in an observed network (\emph{a}) with the weights expected under a null connectivity model (\emph{b}). A modularity matrix is constructed by literally subtracting each edge's expected weight from its observed weight (\emph{c}). Modularity maximization algorithms aim to select clusters such that the stronger-than-expected edges fall within communities. In panels \emph{d} and \emph{e} we show example partitions, one that concentrates strong connections within communities and would be considered of high quality, and another that does not. The detected communities depend on the user's choice of null connectivity model. In \emph{b} we show expected connection weights under a configuration (degree-preserving) null model, which is common in the network neuroscience literature. However, other null models may be more appropriate under certain circumstances. For instance, one can consider a \emph{uniform} model (\emph{f}) in which all connections are assigned an equal weight, a \emph{minimum cost} model (\emph{g}) in which the $m$ shortest connections are formed and their weights assigned inversely proportional distance, a \emph{geometric} model (\emph{h}) where connections and their weights are assigned probabilistically based on distances from one another, a \emph{geometric and degree-preserving} model (\emph{i}) in which both the wiring cost and degree sequence of the original network are preserved, and a \emph{fixed topology} model (\emph{j}) that preserves the exactly configuration of connections as in the original network but where weights are assigned uniformly.} \label{null+models}
	\end{figure*}
	
	The modularity, $Q$, of a partition depends on the structure of our observed network, $W$, as well as the network we would expect by chance -- the matrix $P$ in the modularity expression. This presents a challenge, and in most applications, one of the first instances where the user needs to make a decision. What is the appropriate null model for a network? What is the appropriate null model for a \emph{brain} network? Are there special considerations that need to be taken given how we define our observed network?
	
	Before we address those questions, it is important to explain exactly what the null model does and (does not) represent. In network analyses it is common to compute a summary statistic on a network, e.g. its global efficiency \cite{latora2001efficient}, and compare that value with the distribution of values obtained under some null model, oftentimes one that preserves a set of features from the observed network while randomizing others. The aim of this comparison is to estimate the likelihood that we would obtain the observed efficiency value amongst an ensemble of similar but structureless networks. In the context of modularity maximization, however, the null model serves as an internal comparison and is baked into the modularity function itself. Note, however, that statistical testing can still be carried out after the fact and $Q$ values compared between the observed and random networks.
	
	So what is the appropriate null model for modularity maximization? In practice, most applications use a model that preserves nodes' degrees but otherwise randomizes the placement of links \cite{maslov2002specificity} (Fig.~\ref{null+models}b). In fact, in most software packages, this model is treated as the ``default,'' oftentimes leaving the user no other options. Part of why this model is used so widely is because it is useful -- by preserving nodes' degrees, it ensures that high-degree nodes cannot spuriously inflate the value of $Q$. However, it also has several limitations that make it less appropriate for brain networks.
	
	Notably, the degree-preserving null model -- also known as the Newman-Girvan model or configuration model -- creates random networks through unconstrained edge swaps, resulting in networks with many more long-distance connections. In contrast, real-world anatomical brain networks are constrained by cranial volume and connectional cost (material and metabolic) such that the overwhelming majority of their connections are short-range \cite{stiso2018spatial, kaiser2006nonoptimal, betzel2018specificity, samu2014influence}. In violating this fundamental constraint, degree-preserving null models can be viewed as excessively lenient.
	
	A second related issue concerns dependencies between connections. The connection weights in functional brain networks, because they are calculated using statistical measures like bivariate correlations, are not independent of one another. That is, if we have three nodes $\{ A, B, C \}$ and know the weights of edges $w_{AB}$ and $w_{BC}$, we can place a bound on the weight of edge $w_{AC}$ \cite{zalesky2012use}. However, the networks generated by the edge-swapping model shuffle the placement of edges uniformly and randomly, resulting in configurations that violate statistical limits and are not realizable.
	
	Despite these limitations, the configuration model (and versions of the model made compatible to networks with signed edges \cite{gomez2009analysis}) remains the ``default'' null model for modularity maximization. There exists many alternative options that help resolve these issues (Fig.~\ref{null+models}g-j). For instance, it is straightforward to condition ``edge swaps'' in the configuration model to occur among edges of similar length, which results in a network with approximately the same wiring cost as an observed structural network \cite{betzel2017modular, gollo2018fragility}. In the case of functional connectivity, several approaches exist that address the issue of edge interdependencies. For instance, the uniform null model can be used \cite{bazzi2016community, traag2011narrow}, which tests the null hypothesis that all time series are mutually correlated with one another to some arbitrary extent. Of course, more complicated versions of null models can be imagined, including some based on random matrix theory or \cite{akiki2019determining, almog2019uncovering} others that merge both spatial and correlative information \cite{esfahlani2020space} or preserve topology (the binary pattern of connections) while randomly assigning weights to each edge \cite{bassett2015extraction}.
	
	In general, the choice of null model can impact a partition's quality (larger or smaller values of $Q$) (Fig.~\ref{null+models}d,e). Historically, however, the network neuroscience literature has focused on a single null model and the landscape of alternative models has not been fully explored. Of course, exploring the space of null connectivity models falls beyond the scope of most studies. In those cases, selection of a null model should balance continuity with past work, so as to facilitate clear comparisons, but strive to incorporate more appropriate null models whenever possible.
	
	\subsubsection*{Algorithms for optimizing $Q$}
	
	In general, $Q$ measures the quality of a modular partition -- partitions corresponding to larger values of $Q$ are generally considered better. While this makes it possible to perform comparative analyses of existing, previously-defined partitions, it also opens up the possibility of detecting even better partitions by optimizing $Q$ directly, a procedure referred to as \emph{modularity maximization}.
	
	Although many optimization heuristics have been considered in the past, including simulated annealing \cite{guimera2005functional}, spectral clustering \cite{newman2006finding}, and min-cut algorithms \cite{newman2004fast}, each of which varies both in terms of runtime as well as ability to recover planted communities in synthetic networks, the most popular method is the Louvain algorithm \cite{blondel2008fast}.
	
	The Louvain algorithm is a greedy, two-stage algorithm that scores well on benchmark tests and, in most practical contexts, runs exceptionally fast \cite{lancichinetti2009benchmarks}. The Louvain algorithm is initialized with each node assigned to its own community (although can also begin with a ``seed'' partition); if a network has $N$ nodes, then the number of communities at this stage is equal to $N$. The algorithm then loops over nodes in random order and considers how $Q$ would change if the current node was moved to all of the other communities. The algorithm then accepts the move that results in the greatest $\Delta Q$, subject to the constraint that $\Delta Q > 0$. The loop continues over all nodes and repeats until no node-level moves exist that can improve $Q$. At this point, the algorithm moves to the aggregation stage, in which a meta-network is constructed where each meta-node represents a community from the previous stage and each meta-edge represents that total weight of connections from one meta-node to another. Then, the algorithm returns to the move/merge stage with the meta-network, and attempts to merge meta-nodes into larger communities. In the end, the Louvain algorithm returns community labels for each node, along with the corresponding modularity score, $Q$.
	
	The Louvain algorithm does have limitations, some of which can be addressed with recent extensions. For instance, the Louvain algorithm can lead to poorly-connected communities, meaning that the nodes comprising a community are not directly connected, but only linked to one another by extra-community paths. The recently proposed ``Leiden algorithm'' \cite{traag2019louvain} addresses this and guarantees well-connected communities while preserving the speed and parsimony of the original Louvain algorithm. Nonetheless, the Leiden algorithm was introduced only recently and has yet to be applied in network neuroscience.
	
	\subsubsection*{Stochasticity and consensus clustering}
	
	\begin{tcolorbox}
		
		\textbf{Box 4}. Many community detection algorithms, including the Louvain algorithm for optimizing $Q$, are non-deterministic such that repeated runs can yield dissimilar results. This variability can be resolved using consensus clustering methods.
		
	\end{tcolorbox}
	
	\begin{figure*}[t]
		\centering
		\includegraphics[width=1\textwidth]{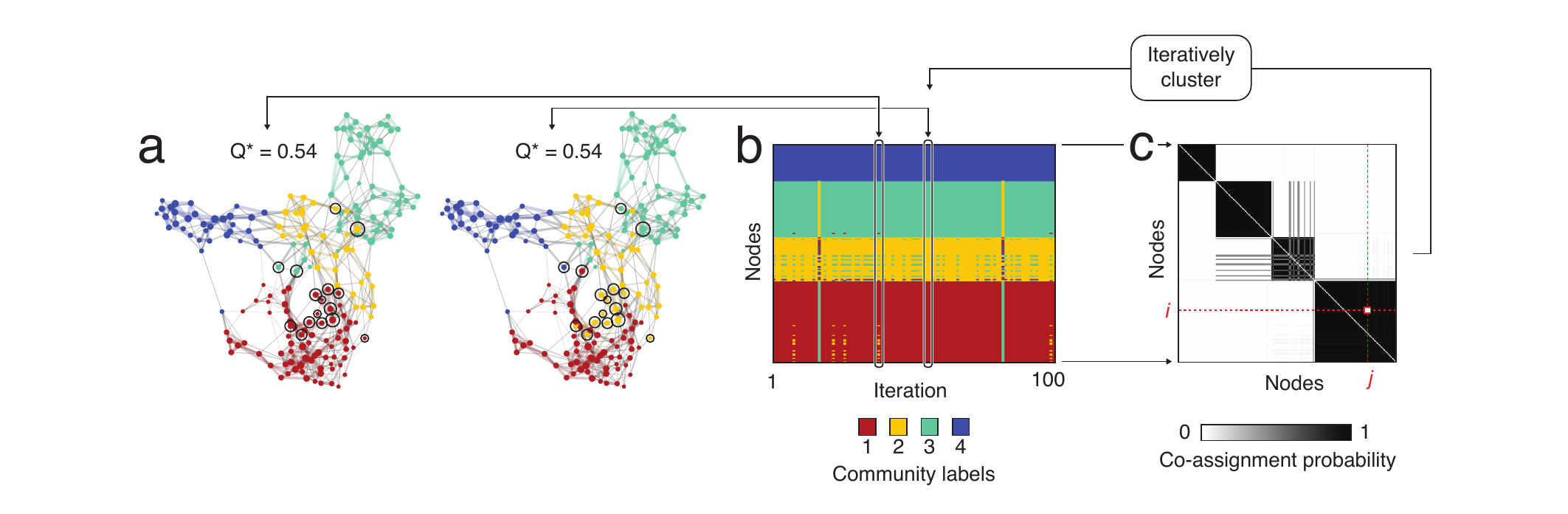}
		\caption{\textbf{Stochasticity and consensus clustering.} The Louvain algorithm is stochastic, such that two runs can produce dissimilar results. Modularity maximization also exhibits a degeneracy of high-quality partitions, meaning that there are likely to be many different partitions of near optimal and equal quality (similar $Q$ value). Consensus clustering is a strategy for combining the outputs of dissimilar partitions into a single, representative partition. (\emph{a}) Two partitions of approximately equal quality ($Q = 0.54$) but different from one another. The dissimilar community assignments are highlighted in with black circles. (\emph{b}) If repeated the Louvain algorithm 100 times, in general we find several different solutions. As part of consensus clustering, we generate a co-assignment or agreement matrix (\emph{c}) from the outputs of multiple Louvain runs. The $\{i,j\}$ element of this matrix counts the fraction of partitions in which nodes $i$ and $j$ were assigned to the same community. This matrix can be clustered using modularity maximization by estimating the expected co-assignment of two nodes to a community under some null model (usually one in which nodes' assignments are uniformly and randomly permuted). Consensus clustering iterates between these two steps: constructing a co-assignment matrix from multiple partitions and clustering it. After a few steps, this algorithm converges so that every run of the optimization algorithm returns an identical partition.} \label{consensus+clustering}
	\end{figure*}
	
	While the Louvain algorithm performs well in practical contexts, it is a greedy and stochastic algorithm, which means that successive runs of the algorithm can (and in general will) result in different estimates of the optimal partition \cite{good2010performance} (Fig.~\ref{consensus+clustering}a). Some studies have embraced this variability, generating ``soft'' partitions of a network by creating a community co-assignment matrix from multiple runs of the algorithm whose elements represent the fraction of those runs in which pairs of nodes were assigned to the same community \cite{hinne2015probabilistic, kenett2020community}.
	
	On the other hand, it is often advantageous to summarize a network's community structure in terms of a definitive, representative partition. There are several strategies for selecting a single partition. One possibility is to simply select the partition with the biggest $Q$ -- after all, $Q$ is a measure of partition quality. While this strategy seems appealing, the near-degeneracy of solutions means that there may exist dissimilar partitions but whose $Q$ values differ by some arbitrarily small amount \cite{good2010performance}. Additionally, given the variability inherent in brain network construction and the noise associated with individual edges and their weights, adopting one partition over another due to a small advantage is $Q$ should be avoided.
	
	A better strategy is to create a representative partition by synthesizing data from multiple partitions, a procedure called consensus clustering . Although there are many methods for performing consensus clustering \cite{strehl2002cluster}, the most popular in network neuroscience involves iteratively clustering a co-assignment matrix until convergence \cite{lancichinetti2012consensus}. The procedure is as follows. Starting with an observed network, a user runs the modularity maximization some fixed number of times, $N_{iter}$, creating an ensemble of high-quality but dissimilar partitions (Fig.~\ref{consensus+clustering}b). From these partitions, one can generate a co-assignment matrix, calculating for nodes $i$ and $j$ the fraction of $N_{iter}$ partitions in which they appeared together in the same community, $D_{ij}$ (Fig.~\ref{consensus+clustering}c).
	
	Then, we set up a new modularity maximization problem by comparing the values in the observed co-assignment matrix with the values we would expect by chance. As before, the chance or null model must be chosen carefully, but one possibility is to compute the expected co-assignment had the community labels in each of the $N_{iter}$ partitions been uniformly and randomly permuted \cite{betzel2020organizing, betzel2018specificity, betzel2017multi} (note that instead of uniform permutations, one can use ``spin'' models to permute nodes' while approximately preserving their relative distances in space \cite{alexander2018testing, markello2020comparing}). This expected level of co-assignment can be calculated directly from the observed co-assignment matrix as the mean of its upper triangle elements. This allows us to write the following modularity expression:
	
	\begin{equation}
	Q_{consensus} = \sum_{ij} [D_{ij} - \langle D \rangle ] \delta (\sigma_i,\sigma_j).
	\end{equation}
	
	\noindent As before, we can use modularity maximization to try and detect the partition that optimizes $Q_{consensus}$. Due to how it is constructed, the co-assignment matrix tends to be ``blockier'' than the original connectivity matrix and detecting its communities is an easier task. This leads to reduced variability in the partitions detected by optimizing $Q_{consensus}$, sometimes to the point that the clustering algorithm consistently returns an identical partition each time. If this occurs, then this partition is considered the consensus partition and the consensus clustering algorithm can stop. If there remains variability in the solutions, then a new co-assignment matrix can be created from the estimated consensus partitions and the algorithm repeats. Note that this procedure diverges slightly from the one proposed in \cite{lancichinetti2012consensus}, where the authors thresholded away elements of the co-assignment matrix below some critical value prior to reclustering.
	
	In general, consensus clustering is a useful tool for resolving variability across multiple estimates of a network's community structure that do not necessarily have to come from the same community detection algorithm. However, it entails some assumptions that a user should be aware of. First, it assumes that there is a single representative set of communities for a network. While this a convenient assumption, recent studies have suggested that the landscape of solutions can be highly degenerate, with multiple near-optimal solutions. Similarly, in extracting consensus clusters, there is an assumption that every node must be assigned to a cluster -- i.e. it is possible to arrive at a consensus about the community to which every node in the network belongs. However, it could be the case (especially if multiple ``good'' partitions exist) that some nodes fundamentally have no clear consensus community assignment \cite{peixoto2020revealing}. In such cases, describing a single optimal partition is not appropriate; an alternative strategy is to describe, statistically, properties of the ensemble of \emph{all} good partitions \cite{peixoto2017nonparametric}.
	
	\subsubsection*{Multi-scale extensions}
	
	\begin{tcolorbox}
		
		\textbf{Box 5}. Modularity maximization can be used to detect communities of different size and number by incorporating a resolution parameter into the modularity expression.
		
	\end{tcolorbox}
	
	\begin{figure*}[t]
		\centering
		\includegraphics[width=1\textwidth]{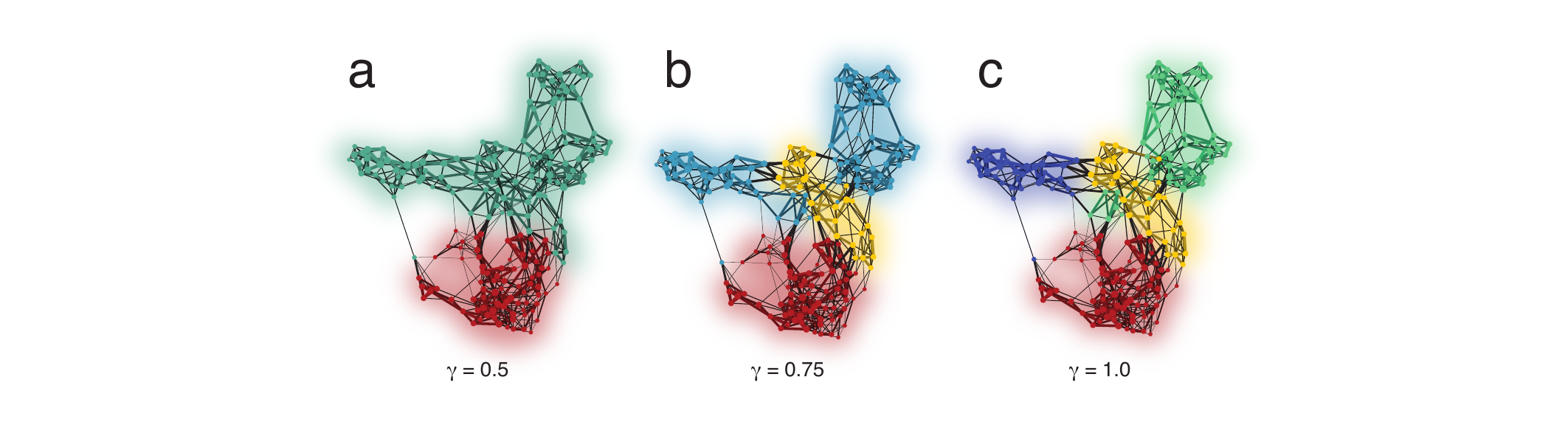}
		\caption{\textbf{Detecting multi-sale community structure using multi-resolution modularity maximization.} Modularity maximization includes a tunable resolution parameter, $\gamma$, to detect communities of different size. In panels \emph{a}, \emph{b}, and \emph{c}, we show three examples of communities in the same network detected at $\gamma = 0.5, 0.75, 1.0$, resulting in 2, 3, and 4 communities .} \label{multiscale+example}
	\end{figure*}
	
	At this point, we have defined the modularity quality function, described some guidelines for selecting its internal null model, reviewed some of the algorithms for optimizing $Q$, and provided a solution for resolving its inherent stochastic nature, yielding (possibly) a single representative consensus partition. At this point, one might expect that the community detection process is over. Unfortunately, this is not the case.
	
	Modularity maximization (and other related methods) can suffer from a limitation known as a ``resolution limit''  \cite{fortunato2007resolution}. The resolution limit means that, for a network of given density, modularity maximization will be unable to detect communities below a certain scale, even if they are unambiguously defined, i.e. perfect cliques. This is because, for certain combinations of null models and networks, especially those that are sparse, the presence of even a single edge between two small communities is considered unexpected. Consequently, it becomes advantageous, in terms of maximizing $Q$, to merge those communities such that unexpectedly strong connection falls within communities, rather than between.
	
	The implication of the resolution limit for modularity maximization is that the communities may not reflect the scale of your network's true and intrinsic community structure and can therefore be misleading (the worst case) or obscure interesting community structure at other scales (a slightly less worse case). Luckily, this issue can be at least partially mitigated through the introduction of a resolution parameter, $\gamma$ to the modularity function \cite{reichardt2006statistical}, so that it reads:
	
	\begin{equation}
	Q(\gamma) = \sum_{ij} [W_{ij} - \gamma \cdot P_{ij}] \delta (\sigma_i, \sigma_j).
	\end{equation}
	
	\noindent Effectively, $\gamma$ scales the importance of the null connectivity model relative to the weights of the observed network. Consequently, when $\gamma$ is small, many connections exceed their expected weights. From the perspective of optimizing $Q$, this makes it advantageous to form large extended communities comprised of many nodes. Conversely, when $\gamma$ is large, only the strongest connections exceed their expected weights, yielding much smaller communities and in greater number.
	
	Incorporating the resolution parameter into modularity maximization makes it possible to change its value and detect communities that might have been undetectable otherwise. It is important to note that, even if one chooses not to incorporate $\gamma$ into their modularity function, it is nonetheless implicitly there, set to its default value of $\gamma = 1$. Additionally, it is worth noting that the particular value of 1 is not privileged or special. In general, it is a good idea to explore the effect of varying $\gamma$ on your detected communities (Fig.~\ref{multiscale+example}a-c).
	
	Additionally, the resolution parameter enables user to detect multi-scale or hierarchical communities -- meaningful groupings of nodes and edges into possibly nested communities of different sizes. This is critical for the study of brain networks where we anticipate ahead of time that the such hierarchical divisions exist, e.g. a bipartition of the brain into task-positive and -negative communities, each of which can be further divided into increasingly functionally specialized sub-systems. Indeed, recent methods have been developed to explicitly address this; given a set of communities sampled at different scales, methods like multi-resolution consensus clustering aim to assemble communities into a natural hierarchy \cite{jeub2018multiresolution}.

	Multi-scale community detection, while useful for circumventing issues related to the resolution limit, nonetheless introduces a free parameter, $\gamma$. While embracing the multi-scale nature of communities may be appealing in some cases \cite{betzel2013multi}, in other applications it is more useful to establish meaningful communities at a single scale corresponding to a specific value of $\gamma$. How might one go about doing this? One possibility is to develop an additional quality function and report communities at the $\gamma$ value where this quality function is optimized \cite{bassett2013robust}. For instance, one could impose communities estimated using one subject's data onto other subjects, and select the $\gamma$ where the modularity induced by the one subject's partitions is maximized. Another possibility is to select the $\gamma$ value where repeated runs of the optimization algorithm yields highly similar partitions, suggesting that at that particular scale, there exists a single well-defined solution \cite{he2018reconfiguration}.
	
	\subsubsection*{Node roles}
	
	\begin{tcolorbox}
		
		\textbf{Box 6}. Once a network's communities have been estimated, measures like participation coefficient and module-degree-zscore can be used to assess node roles within the network, e.g. hub \emph{versus} non-hub.
		
	\end{tcolorbox}
	
	The procedures described above can serve as guideposts for detecting a network's community structure using modularity maximization. However, they are far from definitive; like other data clustering tools, modularity maximization is at times more art than science. Nonetheless, community detection, when done well, can reduce dimensionality and highlight patterns and structure in a network not otherwise evident. Another way that it can be useful is by providing insight into nodes' functional roles within a network.
	
	Two of the best-known measures for ascertaining a node's function role are participation coefficient and module degree-z-score \cite{guimera2005functional}. Each requires that the user know a network's topology and have previously estimated its communities. The participation coefficient, in particular, has proven especially useful in network neuroscience, as it quantifies whether a node's connections are distributed uniformly across many modules (high participation) or concentrated within a relatively small number (low participation) \cite{shine2016dynamics, bertolero2015modular, power2013evidence}. Intuitively, nodes with high participation serve as bridges, binding functionally-specialized modules to one another and regulating the flow of information across their boundaries.
	
	Within-module degree-z-score is a complementary method, and describes how strongly a node is connected to its own module relative the other nodes within its module. Nodes with larger values are generally considered more influential among their neighbors. Together with participation coefficient, within-module degree z-score has been used for identifying hub regions and further classifying them based on whether they are ``connector'' hubs, linking communities to one another, or ``provincial'' hubs that exert local influence over their own module but little influence elsewhere.
	
	Due to its widespread use, participation coefficient has been extended within the network neuroscience community in several ways. These include extending its definition so that it can be applied to networks with signed edge weights, but also modifying the definition to combat some peculiar biases. For instance, \cite{pedersen2020reducing} demonstrated that nodes' participation coefficients are influenced by the connectivity and density of the module to which that node was assigned and introduced a normalization factor to reduce this particular bias.
	
	\subsection*{Limitations}
	
	Community detection can reveal a network's meso-scale structure and, in the case of brains, their system-level architecture. Nonetheless, it has important limitations that users should be wary of. Modularity maximization, specifically, has a number of well-known issues that can limit its applicability or hinder interpretation.

	One of the most common misunderstandings concerns the modularity function and how to interpret $Q$. In most applications, larger values of $Q$ are usually treated as evidence that a network has segregated modules. In general, however, this is not true. Erdos-Renyi networks, where links are formed probabilistically and uniformly, can exhibit high levels of $Q$ due to random fluctuations \cite{guimera2004modularity}. Relatedly, geometric networks, where links are formed based on the proximity of nodes to one another, can also exhibit artificially high levels of modularity, leading users to spuriously detect modules \cite{samu2014influence, betzel2017modular}. In the case of geometric networks, the spuriously high levels of $Q$ are a consequence of using a null model that does not depend on the network's spatial embedding, e.g. a degree-preserving model, as geometric networks exhibit high levels of transitivity due to the underlying metric space in which they are embedded. A simple solution to combat this problem is to incorporate distances into the null model so connections that exist due to spatial effects can be anticipated \cite{expert2011uncovering}.
	
	Another issue associated with community detection, more generally, concerns the lack of a unique solution. Networks are generated according to some process that is a function of its ground truth, ``true'' community structure. Community detection, then, can be thought of as an attempt to invert this process to get at those communities. However, there is not a unique solution to this inversion \cite{peel2017ground}. That is, we can always define some arbitrary function that maps a different set of communities onto our observed network. Accordingly, unless we know the true function (and can therefore invert it), we can never be certain that the communities we detect represent the ground truth. These observations suggest that applications of community detection to real-world networks where the (community-to-network mapping is unknown), should test multiple community detection algorithms and focus on consensus, rather than emphasizing the results of any single approach.
		
	Other concerns include lack of correspondence between ``ground truth'' community structure, e.g. affiliations obtained base on individual's association with an online group, and the community structure detected based on network topology \cite{hric2014community,yang2015defining}. One potential strategy for mitigating this particular issue is to incorporate metadata and vertex annotations directly into the community detection algorithm,  which can be done easily with stochastic blockmodels \cite{newman2016structure}, an approach that has also been applied to functional brain networks \cite{murphy2016explicitly}.
		
	Yet other concerns are specific to the modularity heuristic. For instance, modularity maximization forces every node to have a community assignment, a requirement that, in general, real-world networks are not subjected to (although some methods can reject the hypothesis that a node should be assigned to a community \cite{lancichinetti2011finding}). Relatedly, modularity maximization also forces communities to be non-overlapping, making it impossible for nodes to have multiple community affiliations. In the next section, we introduce some alternative methods that address these limitations.

	\subsection*{Future directions and alternative approaches}
	
	\begin{figure*}[t]
		\centering
		\includegraphics[width=1\textwidth]{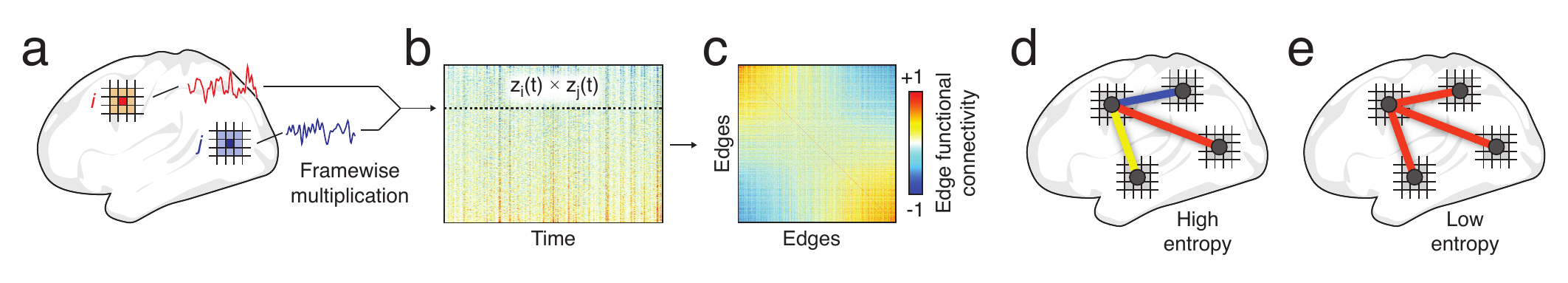}
		\caption{\textbf{Edge functional connectivity and pervasively overlapping community structure.} (\emph{a}) Detecting overlapping communities with edge functional connectivity starts by first generating ``edge time series'' as the element-wise product of z-scored regional time series. (\emph{b}) This procedure is repeated for all pairs of regions. (\emph{c}) Then, the pairwise similarity of edge time series is computed. The result edge-by-edge graph can be clustered. Because edges can take on different community assignments, the nodes upon which they are incident can be associated (\emph{via} their edges) with multiple communities, leading to higher or lower levels of overlap from a node perspective. In panels \emph{d} and \emph{e}, we show examples of a high- and low-overlap scenario from the perspective of the node in the upper left. Following \cite{faskowitz2020edge}, we can quantify the level of community overlap at each node using an entropy measure.} \label{edge+clustering}
	\end{figure*}
	
	Throughout this article we focused on community detection using modularity maximization and highlighted some of the decisions a user must make along the way. Modularity maximization, however, is but one method among a pantheon of methods for detecting a network's community structure and returns communities of a specific character -- non-overlapping, assortative sub-networks. Other methods, however, break free from these assumptions (although they bring along their own sets of assumptions), and can detect communities with vastly different properties. In this section, we briefly review some of the common alternatives to modularity maximization, focusing on those that have been applied to brain network data.
	
	One of the simplest alternatives to ``vanilla'' modularity maximization is to extend it to multi-layer networks \cite{mucha2010community, kivela2014multilayer}. Whereas single-layer modularity maximization can be used to detect communities in one connectivity matrix at a time, multi-layer modularity maximization can detect the community structure of systems composed of multiple connectivity matrices. These matrices could represent estimates of the same network made at different points in time \cite{bassett2011dynamic, finc2020dynamic}, networks from different individuals \cite{betzel2019community}, or network structure under different task conditions \cite{cole2014intrinsic}. The principle advantage of the multi-layer approach is two-fold. First, it resolves the practical issue of how to map communities from one network to another. In general, establishing a correspondence between communities in two networks is not straightforward. There are instances where the correspondence is fundamentally ambiguous, and resolving this ambiguity requires introducing additional heuristics or parameters. With multi-layer modularity maximization, communities are estimated for all layers (networks) simultaneously, such that community labels are preserved across layers, making it trivially easy to assess whether a community in one layer is the same as a community in another. Second, the preservation of community labels makes it possible to assess which communities change their network assignment from one layer to another, a measure referred to as ``flexibility'' \cite{bassett2013robust}, which has been linked to motor and reinforcement learning \cite{bassett2011dynamic, gerraty2018dynamic}, day-to-day variation in affective state \cite{betzel2017positive}, executive function \cite{braun2015dynamic}, language \cite{chai2016functional}, and clinical status \cite{braun2016dynamic}. In general, multi-layer modularity maximization serves as a powerful extension of the traditional single-layer modularity and can be modified and made applicable in any context where community structure for multiple networks (layers) is desired.
	
	Beyond modularity maximization, other popular methods for community include Infomap \cite{rosvall2008maps} and stochastic blockmodels \cite{karrer2011stochastic}. Infomap detects communities based on the trajectory of a random walker moving over a network. Specifically, it aims to reduce an objective function known as the map equation \cite{rosvall2009map}, by assigning each node in the network as short of a ``name'' as possible (in information theoretic terms). If a network has communities that traps the probabilistic flow of a random walker, then instead of assigning each node a unique name, we can assign communities names and reuse the \emph{very} short names for nodes, similar to the way ZIP codes are assigned to cities and street names are reused across different cities. Like modularity maximization, Infomap generates communities that are assortative, internally dense, and externally sparse and has been used widely within the network neuroscience community \cite{power2011functional}.
	
	Yet another method is the stochastic blockmodel. Unlike modularity maximization and Infomap, blockmodels can detect more general classes of community structure, including core-periphery and disassortative (bipartite or multipartite) organization. Blockmodels operate under the assumption of structural equivalence -- whether or not two nodes are connected (and the weight of that connection) depends only on the communities to which those nodes are assigned \cite{karrer2011stochastic} -- essentially grouping nodes together if they have similar connectivity profiles. Although blockmodeling is relatively new in network neuroscience \cite{faskowitz2018weighted, faskowitz2020mapping, betzel2018diversity, moyer2015blockmodels}, it has a long history in the statistical and social sciences \cite{wasserman1994advances, snijders1997estimation}, and continues to be the subject of intense research within network science \cite{peixoto2014hierarchical, peixoto2017nonparametric}.
	
	Modularity maximization, Infomap, and blockmodels generally assume that a network's nodes form non-overlapping communities (although blockmodels can be extended to accommodate overlapping communities \cite{newman2007mixture}). This assumption, however, disagrees with the observation that some brains regions flexibly reconfigure their community assignments over time, maintaining affiliations with multiple overlapping communities. There are, however, many methods that allow for overlap between nodes' community affiliations, although they are relatively uncommon in network neuroscience. For instance, methods like clique percolation \cite{palla2005uncovering} and fuzzy k-means graph clustering \cite{zhang2007identification} generate overlapping estimates of community structure. Others even extend well-established methods like modularity maximization to allow for overlap \cite{nicosia2009extending, najafi2016overlapping}.
	
	Yet another approach for studying community overlap is to shift focus away from a node-centric methods that attempt to cluster a network's nodes and onto edge-centric methods that partition connections into clusters. Because edges are assigned to different communities, the nodes upon which those edges are incident can be associated with multiple communities at the same time. Methods like link similarity \cite{ahn2010link} and line graph clustering \cite{evans2009line} operate this way, transforming node-to-node networks into edge-to-edge networks, which can be clustered using any of the traditional methods. Recently, a novel edge-centric method was proposed for generating overlapping estimates of communities in functional brain networks \cite{faskowitz2020edge, jo2020diversity}. This method works by generating co-fluctuation time series for every pair of nodes as the element-wise product of their z-scored activity \cite{esfahlani2020high} (Fig.~\ref{edge+clustering}a,b). We can calculate \emph{edge functional connectivity} for every pair of time series as their similarity across time, generating a fully weighted and signed edge-by-edge matrix (Fig.~\ref{edge+clustering}c), which can be clustered directly to assign edges to communities, yielding overlapping communities at the level of nodes (Fig.~\ref{edge+clustering}d,e). Whereas other methods for detecting overlapping communities allow for a small number of nodes to participate in multiple communities \cite{gregory2010finding, najafi2016overlapping, li2018collective, palla2005uncovering}, edge communities exhibit pervasive overlap, meaning that every node participates in multiple communities by design.
	
	\section*{Outlook and conclusion}
	
	Community detection can be used to reveal a network's community structure and provide useful insight into its organization and function. However, community detection also prove to be a complicated process that forces a user to define precisely what it means for a network's elements to form a community and make critical decisions about parameters, optimization algorithm, and scale. These decisions make community detection challenging, yet endow users with the freedom to tailor the process to their specific research question.
	
	\section*{Software}
	
	\begin{enumerate}
		\item Brain Connectivity Toolbox (\url{https://sites.google.com/site/bctnet/}):
		\begin{itemize}
			\item \texttt{community\textunderscore louvain.m}: Implements Louvain algorithm for optimizing $Q$.
			\item \texttt{consensus\textunderscore und.m}: Runs threshold-based consensus clustering.
			\item \texttt{participation\textunderscore coef.m} and \texttt{participation\textunderscore coef\textunderscore sign.m}: calculates nodal participation coefficient for networks with positive and signed connections, respectively.
			\item \texttt{module\textunderscore degree\textunderscore zscore.m}: calculates nodal module-degree-zscore.
		\end{itemize}
	\item NetWiki (\url{http://netwiki.amath.unc.edu/GenLouvain/GenLouvain}): Includes function for running a generalized Louvain algorithm (\texttt{genlouvain.m}). This function allows the user to flexibly define their own modularity functions by changing null models and resolution parameters. Includes tutorials for implementing multi-layer and multi-resolution network models.
	\item Network Community Toolbox (requires \texttt{genlouvain.m} function; \url{http://commdetect.weebly.com/}): Suite of functions for optimizing modularity, null models for significance testing, and for further characterizing single communities and entire partitions. Also includes flexibility coefficient.	
	\end{enumerate}
	
	\section*{Acknowledgements}
	
	Thanks to Haily Merritt, Jacob Tanner, and Joshua Faskowitz for reading an early version of this manuscript and providing feedback.
	
	\bibliography{biblio_communities}

\end{document}